\documentclass{article}

\usepackage{arxiv}

\usepackage[utf8]{inputenc} 
\usepackage[T1]{fontenc}    
\usepackage{hyperref}       
\usepackage{url}            
\usepackage{booktabs}       
\usepackage{amsfonts}       
\usepackage{nicefrac}       
\usepackage{microtype}      
\usepackage{lipsum}		
\usepackage{graphicx}
\usepackage{doi}
\usepackage{siunitx}
\usepackage{soul}
\renewcommand\hl[1]{#1}

\title{Quantifying the trade-off between stiffness and permeability in hydrogels}

\author{ \href{https://orcid.org/0000-0002-8834-1664}{\includegraphics[scale=0.06]{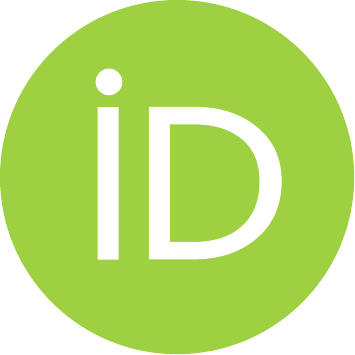}\hspace{1mm}Yiwei Gao} \\
	Department of Mechanical Engineering\\
	University of Nevada, Las Vegas\\
	Las Vegas, NV 89052 \\
	\texttt{yiwei.gao@unlv.edu} \\
	\And
	\href{https://orcid.org/0000-0001-5857-0055}{\includegraphics[scale=0.06]{orcid.pdf}\hspace{1mm}H. Jeremy Cho}\thanks{\url{https://dakine.sites.unlv.edu}} \\
	Department of Mechanical Engineering\\
	University of Nevada, Las Vegas\\
	Las Vegas, NV 89052 \\
	\texttt{jeremy.cho@unlv.edu} \\
}

\renewcommand{\shorttitle}{Quantifying the trade-off between stiffness and permeability in hydrogels}

\hypersetup{
pdftitle={A template for the arxiv style},
pdfauthor={Yiwei Gao, H. Jeremy Cho},
}

\begin{document}
\maketitle

\begin{abstract}
	Hydrogels have a distinct combination of mechanical and water-transport behaviors. As hydrogels stiffen, they become less permeable. Here, we combine semi-dilute polymer theory with the Kozeny-Carman equation to develop a simple scaling law describing the relationship between hydraulic permeability and mechanical stiffness. We find that these properties are dictated by the polymer strand spacing, explained via analogy of a bowl of noodle soup. We find a remarkably close agreement between our scaling law and experimental results across four different polymer families with varied crosslinkings.
\end{abstract}


Hydrogels are polymer networks with an interconnected, water-filled porous structure. Water can be absorbed within these pores or permeate through them. The polymer network provides mechanical structure to  hydrogels. This leads to hydrogels having a distinct combination of mechanical and water-transport behaviors, making them advantageous for a wide variety of applications. Much of the studies that involve mechanics and transport are application-focused and either study the mechanics in detail (e.g., stimuli-responsive water flow materials, soft robotics, and artificial tissues \cite{tan2019application,zhou2019highly,subramani2020influence,liu2020programmable,lin2016stretchable,jiang2011pva}) or transport behavior (e.g., contact lenses, drug delivery, wastewater purification, and solar distillation \cite{efron2007oxygen,kim2008extended,zhao2019novel,sinha2019advances,lu2021high,van2018hydrogel,jones2008characterization}). However, no study has provided a concrete coupling of mechanics and water transport from a fundamental molecular perspective.

We believe that crosslinking provides an important clue as to how molecular structure controls permeability and stiffness. Previous studies have heavily relied on varying crosslinking density to modify stiffness, permeability, as well as a multitude of other properties.  Stiffness-focused studies \cite{lin2015influence,pilipchuk2013influence,gao2021scaling} have found that  higher crosslinking densities generally result in stiffer gels. Transport-focused studies have found that crosslinking can modify gas or solute permeability \cite{ju2010characterization,peng2012ion,pozuelo2014oxygen,efron2007oxygen,liu2008gas} where higher crosslinking densities generally result in less permeable gels \cite{matsuyama1997analysis,yazdi2020hydrogel,sagle2009peg,tavera2018characterization}. Using crosslinking as an experimental variable, how molecular structure is coupled with mechanics and transport is the central question that we seek to answer.

To answer this question, we adopt de Gennes' semi-dilute polymer solution theory \cite{de1979scaling} used in our previous work \cite{gao2021scaling} where we characterized changes in stiffness, osmotic pressure, and swelling by varying crosslinking density. The semi-dilute theory is a simple, elegant, and experimentally proven description that produces results that are indistinguishable from the more thorough, but complex Flory-Rehner theory \cite{vandersman2015biopolymer}. The semi-dilute theory provides an illustrative molecular perspective of the most essential part of polymer mixtures. With this perspective, the gel is considered to be a solution of polymer chains that are long and the spacing between the chains, $\xi$, is what ultimately dictates the properties of the gel. This spacing between polymer chains is related to the polymer volume fraction as $\xi\propto \phi_\text{poly}^{-3/4}$ where $\phi_\text{poly}\equiv V_\text{poly}/V$. Using this theory in our previous work, we found that the bulk modulus ultimately depends on the polymer volume fraction---or alternatively the amount of gel swelling, which is inversely proportional to the polymer volume fraction.  Thus, from a molecular perspective, changing the crosslinker ratio does not change the stiffness; rather, changing the crosslinker ratio modifies the polymer volume fraction and spacing between polymer chains, which ultimately affects the elastic bulk modulus. This modification of the bulk modulus also equivalently changes the osmotic pressure, $K \sim \Pi$ \cite{de1979scaling,gao2021scaling}. In a semi-dilute system, the precise relationship (des Cloiseaux law) between stiffness, osmotic pressure, and polymer volume fraction is
\begin{equation}
K \sim \Pi \sim kTa^{-21/4}v^{3/4}\phi_\text{poly}^{9/4}  \label{Eq.1full}
\end{equation}
where $k$ is the Boltzmann constant, $T$ is absolute temperature, $a$ is the monomer size, and $v$ is the excluded volume of the
monomer that depends on the Flory interaction parameter, $\chi$ \cite{gao2021scaling}. From Eq.~\ref{Eq.1full}, we highlight how stiffness scales with polymer volume fraction:
\begin{equation}
    K\propto \phi_\text{poly}^{9/4}, \label{Eq.1}
\end{equation} 
quantifying the effect that gels soften ($K$ decreases) as they swell with water ($\phi_\text{poly}$ decreases). We and others previously verified this $9/4$ scaling \cite{gao2021scaling,ZRINYI19871139,obhukhov,bellpeppas}, validating the semi-dilute description for hydrogels.

Inspired by the fact that $\phi_\text{poly}$ is an important controlling parameter of gels, we investigate whether $\phi_\text{poly}$ also controls the water transport behavior of hydrogels as quantified by the hydraulic permeability. Intuitively, we should expect that by increasing crosslinker ratio, the spacing between polymer chains, $\xi$, decreases, causing the polymer volume fraction to increase. As a result, the polymeric network becomes more constrained, with less porous space \cite{da2007effect,ben2018comparative,collins2011morphology,bryant2004crosslinking}. Here, we investigate whether this more constrained polymer network increases the difficulty to flow water through it as quantified by the Darcy hydraulic permeability, $\kappa$: 
\begin{equation}
    \kappa = \frac{\mu u }{\nabla P}, \label{Eq.2}
\end{equation} 
where $\mu$ is the dynamic viscosity of water, $u$ is the volumetric flow flux, and $\nabla P$ is the hydraulic (pore) pressure gradient. To test whether higher $\phi_\text{poly}$ indeed results in lower permeability, we vary crosslinker amount as a means to adjust the polymer volume fraction while keeping other molecular properties unchanged. That is, in Eq.~\ref{Eq.1full}, the monomer size, $a$, and excluded volume, $v$, do not change with crosslinker amount---only $\phi_\text{poly}$ changes.

\begin{figure}[htb]
    \includegraphics[width=0.5\columnwidth]{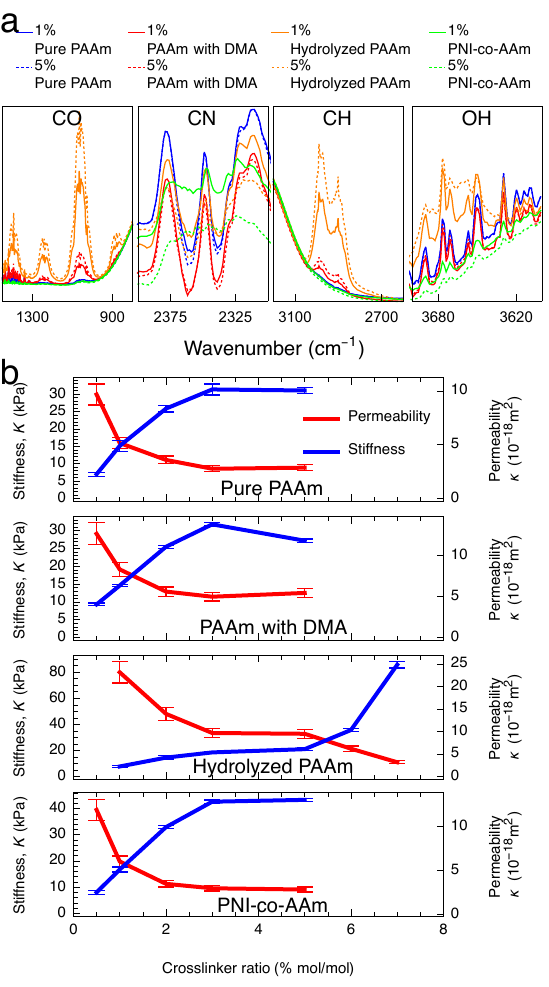}
    \centering
    \caption{a) Significant changes in FTIR spectra are only observed across polymer families and not by varied crosslinker ratio. (Complete FTIR test results of all samples are shown in SI, Fig.S5.) b) In all four hydrogel families, increasing crosslinker ratio generally increases stiffness and decreases hydraulic permeability with a few polymer-family-specific exceptions. Crosslinker ratios above \SI{5}{\percent} are not tested for some hydrogels due to high sample brittleness precluding accurate stiffness characterization.}
    \label{Fig 2}
\end{figure}

 We verify that crosslinker amount does not affect other molecular properties except $\phi_\text{poly}$ through osmotic pressure testing as we and others have previously shown \cite{gao2021scaling,li2012experimental}. We find that osmotic pressure does not change with crosslinker amount, confirming the invariance on $a$ and $v$. As such, hydrogels that differ only by crosslinking can be viewed as being a part of the same \textit{polymer family}. To provide a comprehensive understanding, in this study, we consider 21 different hydrogels spanning across four different polymer families: (1) pure polyacrylamide (PAAm) hydrogels, (2) hydrolyzed PAAm hydrogels \cite{saber2012uv,aalaie2007preparation}, (3) PAAm hydrogels with N,N-dimethylacrylamide (DMA) as a filler \cite{skelton2013biomimetic,kuru2007preparation}, and (4) PAAm hydrogels copolymerized with N-Isopropylacrylamide (PNI-co-AAm) \cite{manjula2013preparation,yang2020preparation}. Within each family, we have five to six different hydrogels differing by crosslinking amount where we use  N,N’-methylene(bis)acrylamide (MBA) as a crosslinker. The molar crosslinker-to-monomer ratios span \SIrange{0.5}{7}{\percent}. Fourier-transform infrared (FTIR) spectroscopy analysis confirms that significant changes in spectra are observed across different families, but insignificant changes are observed across different crosslinker ratios within the same families (Fig.~1a), consistent with our previous work \cite{gao2021scaling}.

\begin{figure*} [t]
    \centering
    \includegraphics[width=0.9\textwidth]{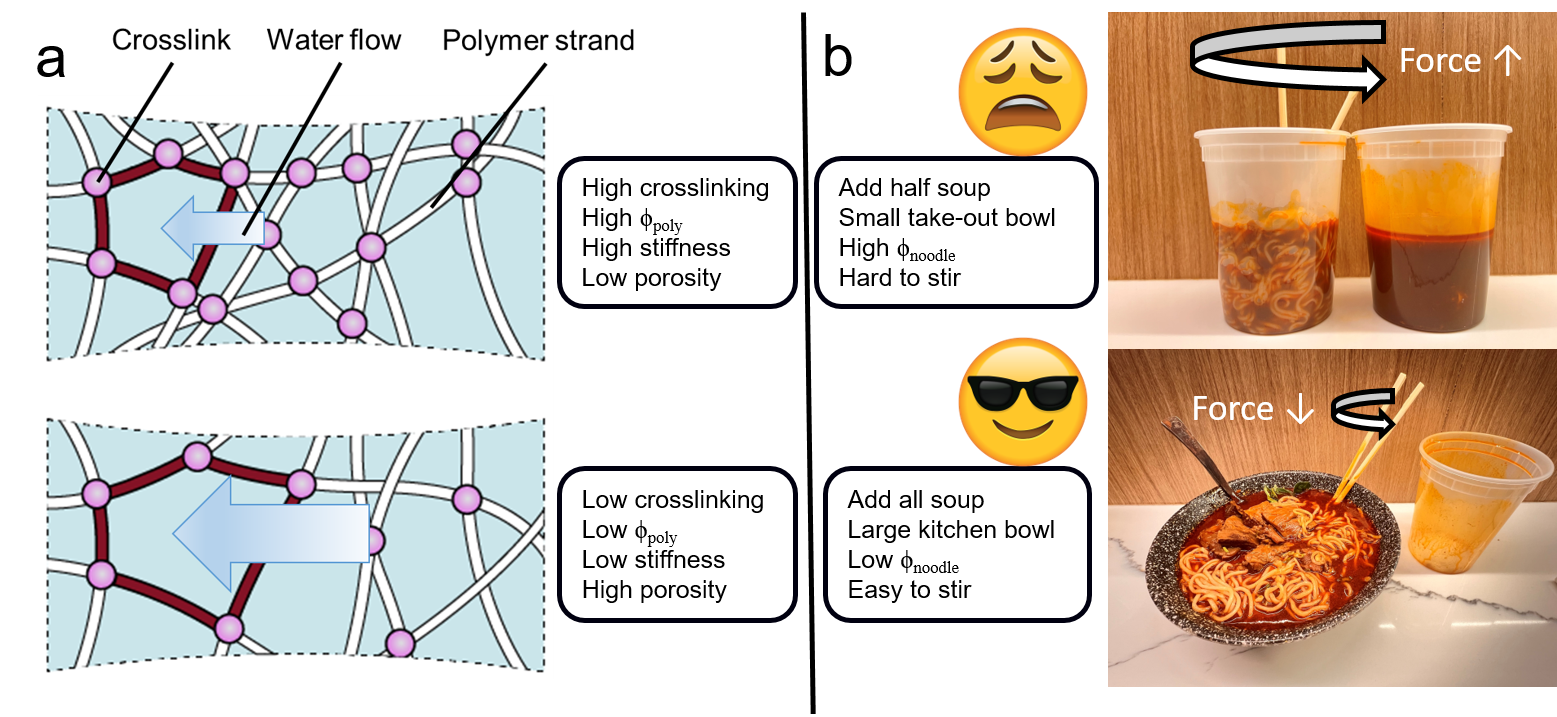}
    \caption{a) A scheme of polymeric structure of hydrogels with low and high crosslinking illustrates the differences in mechanical stiffness and water transport. As the crosslinking increases, the network densifies, resulting in higher stiffness and less porous space, reducing permeability. b) Analogously, a take-out container of noodle soup that is unable to hold all of the soup has a high $\phi_\text{noodle}$, resulting in a mixture that is harder to stir and mix in any desired condiments.}
    \label{Fig 1}
\end{figure*}
To test whether permeability is reduced in hydrogels of higher polymer volume fraction, we perform water transport experiments using a custom-built permeability cell controlled by a microfluidic pressure regulator (Elveflow OB1) and flow rate sensor (see SI, Fig~S1). Using Darcy’s law (Eq.~\ref{Eq.2}), we find that in the limit of a small pressure difference, $\Delta P$, applied across the gel, the experimentally measurable permeability is
\begin{equation}
    \kappa \approx \frac{\mu Q L}{A \Delta P} \label{Eq.2.experimental}
\end{equation} 
where $Q$ is the volumetric flow rate, $L$ is the sample thickness, and $A$ is the cross-sectional flow area. Here, we measure $Q$ and control $\Delta P$ while $\mu$, $A$, and $L$ are fixed, measurable constants. If $\Delta P$ is appropriately low, then $\kappa$ would be a constant property and $Q$ would be \textit{linear} with $\Delta P$. On the other hand, if $\Delta P$ is too high, $L$ would decrease due to poroelastic compression of the gel and $\kappa$ would presumably decrease due to densification of the porous network, leading to a \textit{nonlinear} relationship between $Q$ and $\Delta P$. Informed by previous work on hydrogels under confined loads \cite{louf2021underpressure}, the threshold pressure above which these nonlinear effects take hold is related to the bulk modulus of the hydrogel, $K$.  Varying $\Delta P/K$ in the range of 0.1--2, we find $Q$ is highly linear with $\Delta P$ when $\Delta P/K$ is in the range of 0.5--1 (lower pressures introduce measurement uncertainties); thus, we are confident in our value of $\kappa$ within an interval of $\pm \SI{2}\percent$ (see SI, Fig.~S2). Therefore, to optimize measurement fidelity and minimize poroelastic compression effects, all gels are tested by setting $\Delta P/K = 0.7$.

As expected, the permeability of the tested hydrogels decreases as the crosslinker ratio generally increases from \SIrange{0.5}{7}{\percent} (Fig.~1b, red). However, there are a few notable complications. Increasing crosslinker ratio between \SIrange{3}{5}{\percent} results in very little change (plateau-like trend) or even an increase in permeability (PAAm with DMA). As shown in the hydrolyzed PAAm samples above \SI{5}{\percent}, the permeability sharply decreases in contrast to the plateau-like behavior at slightly lower crosslinker ratios. Thus, the relationship between permeability and crosslinker ratio is highly polymer-family-specific and would likely need a complicated model to describe accurately. As discussed earlier, we believe that polymer volume fraction more directly controls permeability as it does with stiffness. If this were the case, then any changes in permeability should have corresponding changes in stiffness. Therefore, we also measure the stiffness of the tested samples using a custom-built indentation tester (see SI, Fig.~S3) in accordance with previous testing procedures \cite{gao2021scaling}. Indeed, the stiffness of the hydrogels increases as crosslinker ratio increases (Fig.~1b, blue). Similar to our measurements with permeability, the dependence on crosslinker ratio is highly polymer-family-specific. However, as expected, any changes in permeability have inversely proportional changes in stiffness. This coupling between permeability and stiffness is highlighted by the fact that while increasing crosslinker ratio from \SI{3}{\percent} to \SI{5}{\percent} for PAAm with DMA results in an anomalous increase in permeability, this is complemented by an anomalous decrease in stiffness. Furthermore, any plateau-like trends in permeability also exist in stiffness. This suggests that both permeability and stiffness are governed by the more fundamental polymer volume fraction and spacing between polymer chains as opposed to the crosslinker ratio (Fig.~2a).

To understand how an inverse relationship between permeability and stiffness can result from a change in polymer volume fraction, we invite the reader to entertain a thought experiment on a bowl of noodles in soup (Fig. 2b). If the volume fraction of noodles, $\phi_\text{noodle}$, is high, then it is difficult to stir and mix in any condiments into the container---analogous to high stiffness and low permeability with a high $\phi_\text{poly}$ gel (Fig.~2a). This scenario is often encountered when an insufficiently large take-out noodle container is unable to contain all of the soup (high $\phi_\text{noodle}$), making it difficult to mix around, incorporate any condiments, and consume. Whereas, if the noodles and all of the soup were combined together in a sufficiently large bowl (low $\phi_\text{noodle}$), then it would be easier to stir and mix in any condiments, facilitating easier consumption.

Whether it be noodles or polymer chains, we can quantify the relationship between volume fraction and permeability using a model that depends on porosity. The Kozeny-Carman equation is a widely used model in geology that can serve this purpose. While it was originally derived for a packed bed of solid particles, it has been used to describe permeability through gel materials \cite{nakao1979characteristics,zaidi2005experimental,iritani2006compression,isobe2018poroelasticity}. Specifically, the Kozeny-Carman equation quantifies the hydraulic permeability as 
\begin{equation}
    \kappa = C \frac{\epsilon^{3}}{(1-\epsilon)^{2}}, \label{Eq.3}
\end{equation}
where $C$ is a constant and $\epsilon$ is the porosity (porous volume fraction). Assuming the hydrogel polymer matrix is effectively a packed bed of polymer chains, the porosity is related to the polymer volume fraction as $\epsilon = {1}-\phi_\text{poly}$. Thus, the hydraulic permeability, in terms of the polymer volume fraction, is
\begin{equation}
    \kappa \propto \frac{({1}-\phi_\text{poly})^{3}}{\phi_\text{poly}^{2}} = \frac{1}{\phi_\text{poly}^2}-\frac{3}{\phi_\text{poly}}+3-\phi_\text{poly}
    \text{.} \label{Eq.expandedkappa}
\end{equation}

\begin{figure}[htb]
    \includegraphics[width=0.5\columnwidth]{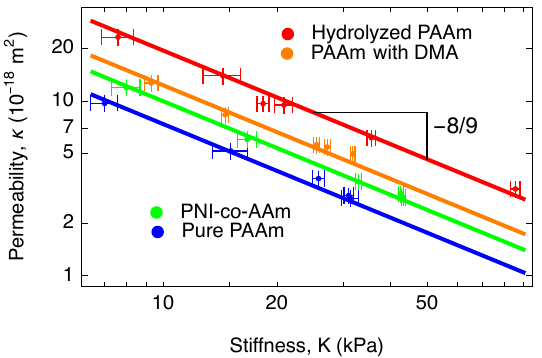}
    \centering
    \caption{Plotting permeability and stiffness of 21 hydrogels with varying crosslinker ratio across four polymer families shows a close agreement with Eq.~\ref{Eq.scalinglaw} where the log-log slope is $-8/9$.}
    \label{Fig 3}
\end{figure}

In the limit of small $\phi_\text{poly}$---applicable to highly swollen gels---the magnitude of the leading-order term, $1/\phi_\text{poly}^2$, is greater than the other terms. Thus, for small $\phi_\text{poly}$ the permeability can be approximated by dropping other terms as
\begin{equation}
    \kappa \propto \phi_\text{poly}^{-2}. \label{Eq.4}
\end{equation}
If this approximation is valid, then the slope of $\kappa$ using Eq.~\ref{Eq.expandedkappa} when plotted on a log-log scale should be similar to $-2$ as this is the exponent of the simplified scaling relationship in Eq.~\ref{Eq.4}. Indeed, as long as $\phi_\text{poly} < {0.05}$, the similarity in log-log slope between Eq.~\ref{Eq.expandedkappa} and Eq.~\ref{Eq.4} is within \SI{8}{\percent} (see SI, Fig.~S8). Since, for many swollen hydrogels, $\phi_\text{poly}$ is often much smaller than $0.05$ \cite{zhang2006surprising,lv2019enhanced,kim2003swelling,kumar2010synthesis}, the relationship in Eq.~\ref{Eq.4} is a very reasonable simplification, providing single-digit-percentage uncertainty or less.

Combining the scaling relationships between stiffness to $\phi_\text{poly}$ (Eq.~\ref{Eq.1}) and permeability to $\phi_\text{poly}$ (Eq.~\ref{Eq.4}),  and subsequently eliminating $\phi_\text{poly}$, we obtain a scaling law that directly relates stiffness, $K$, to the hydraulic permeability, $\kappa$:
\begin{equation}
    \kappa \propto K^{-8/9}. \label{Eq.scalinglaw}
\end{equation}
This scaling law quantifies the inverse relationship between permeability and stiffness by coupling the semi-dilute polymer description of modulus and the Kozeny-Carman description of hydraulic permeability. This relationship should hold for any hydrogel that varies only by polymer volume fraction, such as if crosslinking were modified. To confirm this simple scaling law, we plot the permeability against the stiffness for the 21 hydrogels across four different polymer families with varied crosslinking from \SIrange{0.5}{7}{\percent} tested earlier on log-log scaling (Fig.~3). In accordance with Eq.~\ref{Eq.scalinglaw}, each polymer family has data points corresponding to different crosslinker ratios that lie on a slope of ${-8/9}$. The close agreement within \SI{3.7}{\percent} across a wide range of hydrogel crosslinker ratios provides strong validation of the scaling law (Eq.~8). We highlight the fact that for hydrolyzed PAAm gels, we are able to obtain more than a decade range of stiffness and find that the scaling still holds. Thus, our scaling law allows one to predict the permeability of gels of any arbitrary $\phi_\text{poly}$ whether they be controlled by crosslinking or by other means.

In conclusion, our work confirms that combining de Gennes’ semi-dilute polymer theory and the Kozeny-Carman equation describing hydraulic permeability provides an accurate scaling law relating permeability to the stiffness of hydrogels. A key insight gained from this work is that both permeability and stiffness are ultimately dictated by the polymer volume fraction, which is inherently tied to the spacing between polymer chains. Thus, through this polymer volume fraction, we provide a new, fundamental perspective of how molecular structure affects the coupling of mechanics and water transport. Future studies could investigate whether our scaling law can be applied to hydrogels with high entanglement \cite{kim2021fracture,chen2019entanglement} or sliding crosslinkers \cite{gong2010double,du2020highly,shi2020ultrastrong} to alter polymer volume fraction. We anticipate that our work will guide the informed synthesis of hydrogels tuned by crosslinker amount for applications in a wide variety of fields. Furthermore, the molecular persepective of this work can help uncover how polymer strand spacing controls other aspects, behaviors, and properties of hydrogels.

\section*{Acknowledgements}
It is a pleasure to acknowledge Brandon Ortiz and Ryan Phung for helpful discussions and assistance in experiments as well as Nicholas K. K. Chai for the design and build of custom indentation tester. We also thank Suraj V. Pochampally and Jaeyun Moon for assistance with FTIR spectroscopy. This work was supported by the University of Nevada, Las Vegas through
start-up funds, the Faculty Opportunity Award, the Top Tier
Doctoral Graduate Research Assistantship program.
\clearpage

\renewcommand\thefigure{S.\arabic{figure}}    
\setcounter{figure}{0}    
\section*{Supplementary Information}
\subsection*{Chemicals}
The chemicals used in the preparation of hydrogels are listed below:

Acrylamide (AAm)

N,N' - Methylenebis(acrylamide) (MBA)

Ammonium Persulfate (APS)

N,N,N',N' - teramethylethane - 1,2 - dimine (TEMED)

N,N - Dimethylacrylamide (DMA)

N-Isopropylacrylamide (NIPAAm)

All chemicals used in this paper are purchasd from Sigma-Aldrich Co.

\subsection*{Hydrogel Preparation}

All hydrogels were prepared from aqueous stock solutions of the following chemicals: N,N’-methylene(bis)acrylamide (MBA), N,N-Dimethylacrylamide (DMA), ammonium persulfate (APS), and tetramethylethylenediamine (TEMED) at concentrations of \SI{0.1084}{\gram}/\SI{10}{\milli\liter}, \SI{2.6}{\milli\liter}/\SI{10}{\milli\liter}, \SI{0.08}{\gram}/\SI{10}{\milli\liter} and \SI{0.25}{\milli\liter}/\SI{10}{\milli\liter}, respectively. The base acrylamide (AAm) monomer was used in its pure powder form. By mixing different amounts of these chemicals, polymers were spontaneously synthesized. During this process, APS served as an initiator, TEMED as an accelerator, and MBA as a crosslinker.

In all hydrogels, we started with \SI{0.5}{\gram} of AAm monomer, \SI{1}{\milli\liter} of TEMED solution and \SI{1}{\milli\liter} of APS solution. Amounts of MBA solution waws varied to achieve the target crosslinker ratio (MBA/AAm, mol/mol) ranging from \SIrange{0.5}{7}{\percent}.

For hydrogels with DMA, \SI{10}{\percent} (DMA/AAm mol/mol) was added. Then, the solution was vortex mixed for approximately one minute and subsequently rested at room temperature (\SI{24}{\celsius}) for 24 hours. For samples that were hydrolyzed, we immersed the samples in \SI{1}{\mole\per\liter} sodium hydroxide for 30 minutes before. For PNI-co-AAm hydrogels, \SI{10}{\percent\mole} (of AAm) of NIPAAm was added before adding crosslinker, accelerator and initiator. The ratio of Total Water/AAm was fixed to be 1000 (mL/mol).

Finally, the samples were rinsed everyday and immersed in DI water for one week to remove unreacted chemicals and equilibrate them to the wet state.

\clearpage

\begin{figure}[htb]
    \centering
    \includegraphics[width=0.75\linewidth]{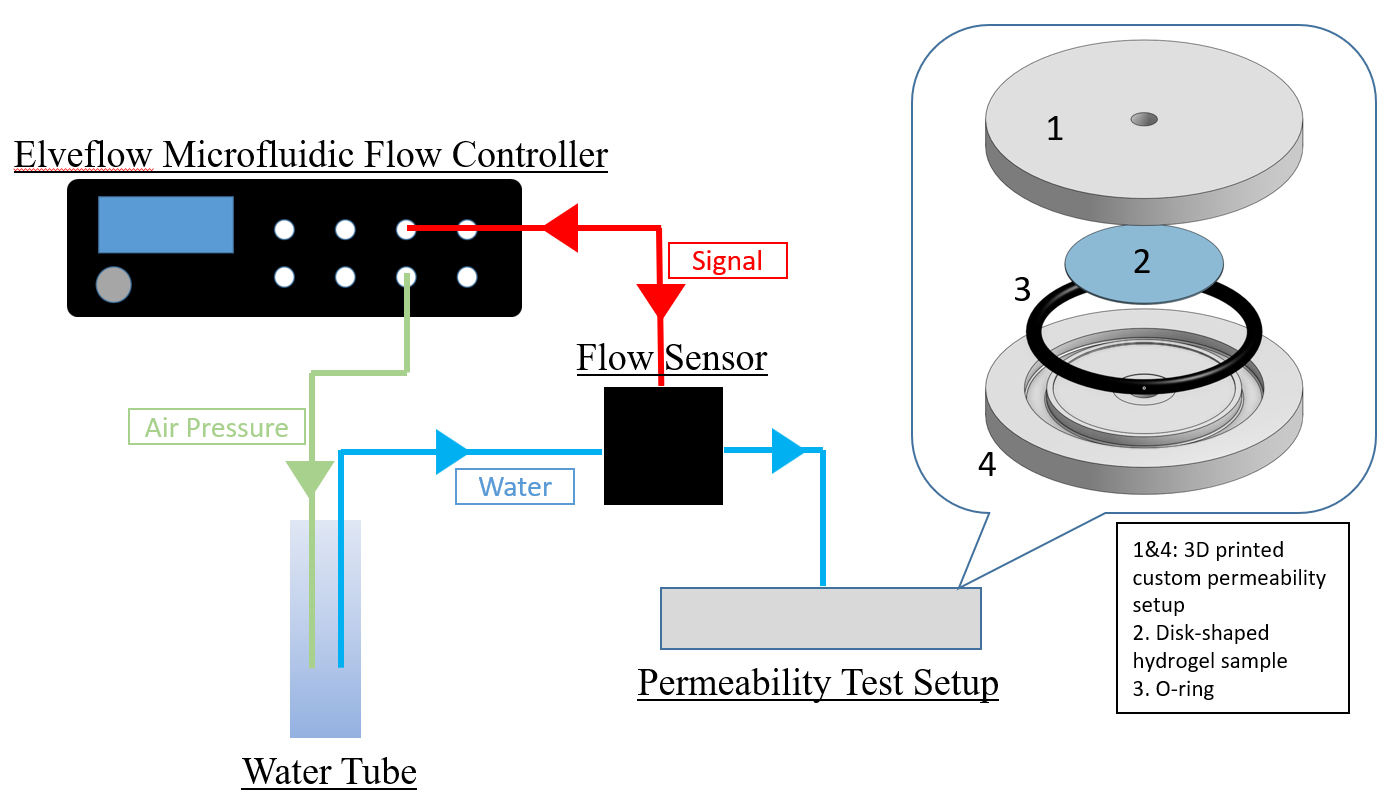}
    \caption{A scheme of the custom-built permeability tester. The custom permeability setup was designed and printed in our lab. With the setup, the gel thickness, $L$, and cross-sectional area, $A$, (Eq.~4) can be fixed at certain values. To ensure the sealing of the setup, an o-ring was applied between top and bottom parts. After the sample was inserted in the setup, two clamps were used to keep the setup being tight through the testing. $\Delta P$ (Eq.~4) was controlled by the Elveflow Microfluidic Flow Controller and provided pressure in the water tube. Water in the water tube was pushed out and ran through the flow sensor, where instant reading of the volumetric flow rate, $Q$ (Eq.~4) could be sent back to the controller for an accurate adjustment. Water flow with targeting pressure flew through the sample in the permeability setup and a real-time $Q$ was read by the sensor.}
    \label{SM Figure 3}
\end{figure}

\clearpage
\begin{figure}
    \centering
    \includegraphics[width=0.75\linewidth]{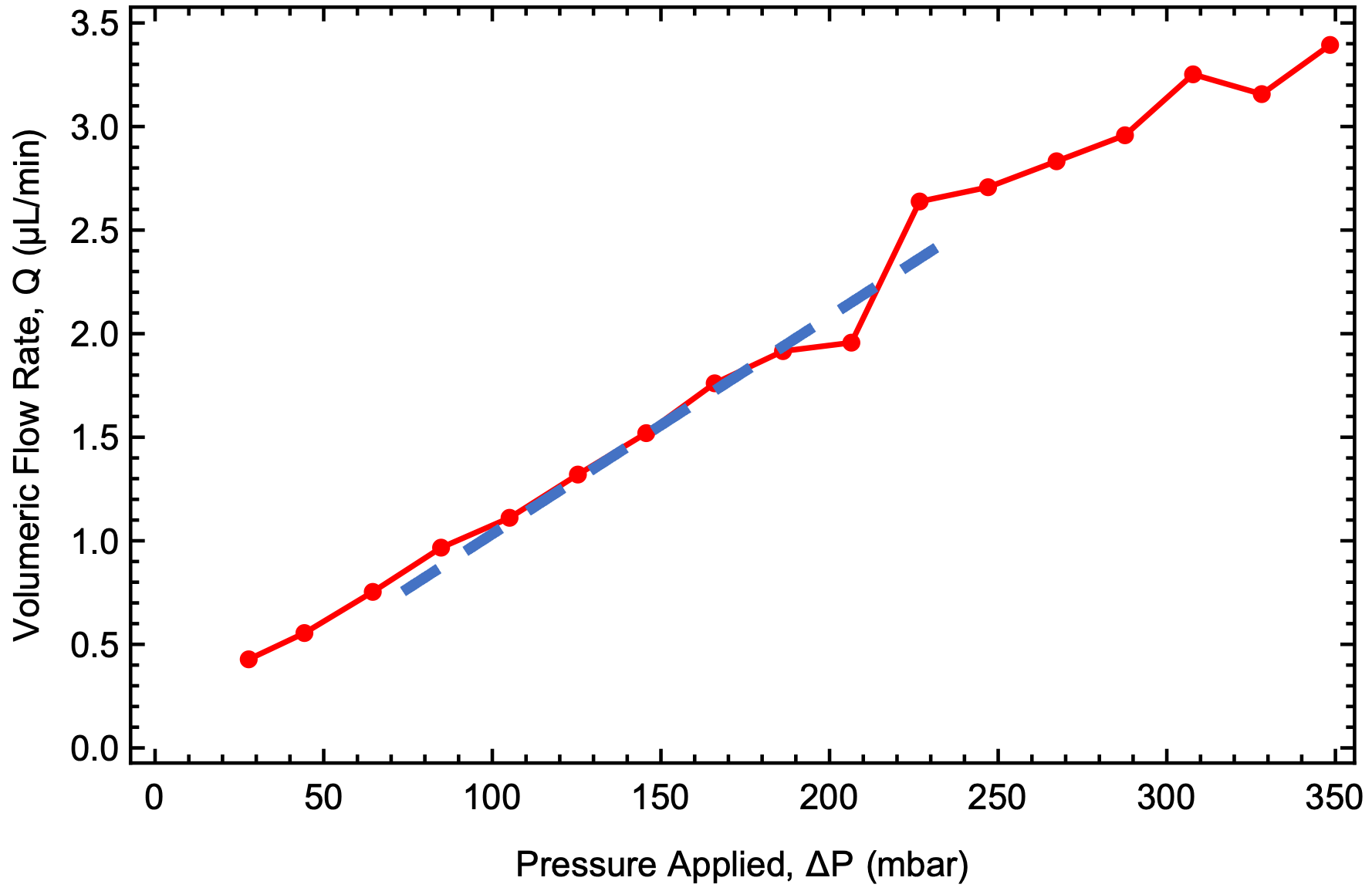}
    \caption{$K$ of the tested sample was \mbox{\SI{18.4} kPa (\SI{184}{\milli\bar})}. We varied pressure and found that the slope of $Q/\Delta P$ \hl{was nearly linear} when $\Delta P/K$ was in the range of 0.5--1 (lower pressures introduce measurement uncertainties, within an interval of \mbox{$\pm \SI{2}\percent$)}; therefore, we set $\Delta P$ such that \mbox{$\Delta P/K$ = 0.7} for all our permeability tests of all samples.}
    \label{SM Figure 5}
\end{figure}

\clearpage

\begin{figure*} [htb]
    \centering 
    \includegraphics[width=0.8\linewidth]{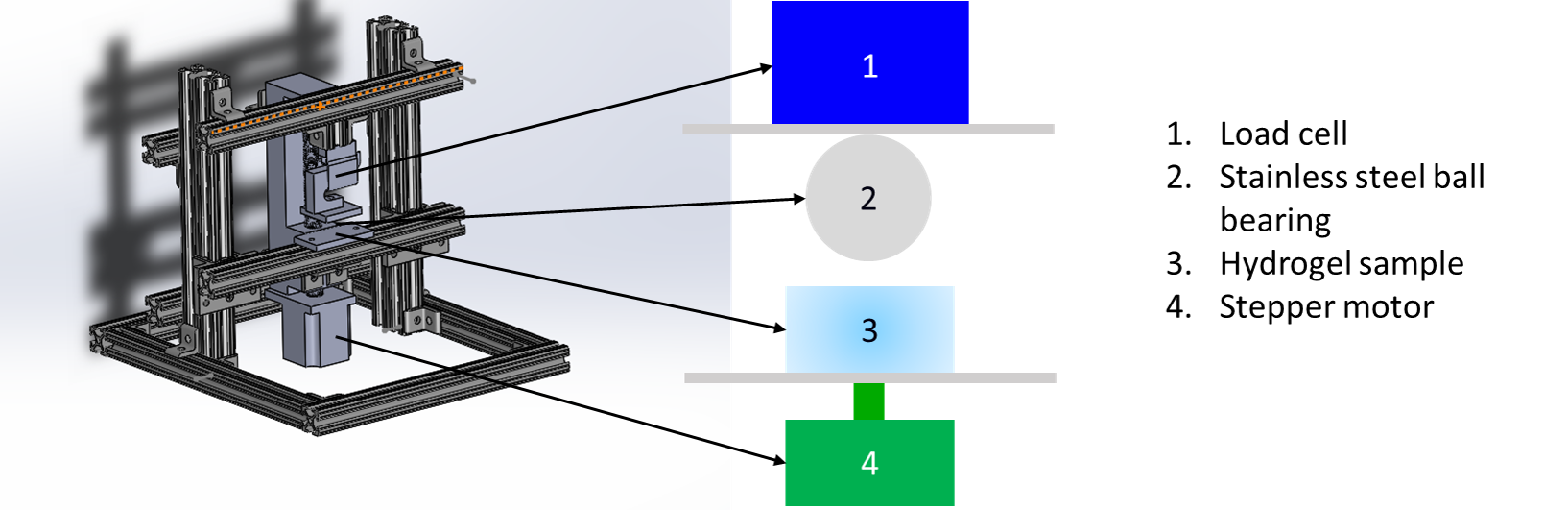}
    \caption{We used a custom-built indentation tester to perform these measurements. Samples were prepared in a cylindrical shape and oriented such that a flat surface was indented. All tests are completed within 15 minutes to ensure minimal weight loss from de-swelling to the ambient environment. To confirm this minimal weight loss, we ensured that the weights of the samples before and after indentation tests were less than {\SI{1}{\percent}}. Displacement speeds ranged from \SIrange{5}{10}{\milli\meter\per\minute}; slower or faster speeds did not affect the force-displacement curves, indicating the sample behaved quasi-statically and quasi-elastically, away from dynamic drainage and viscoelastic effects.}
    \label{SM Figure 1}
\end{figure*}

\clearpage

\begin{figure}
    \centering
    \includegraphics[width=0.75\linewidth]{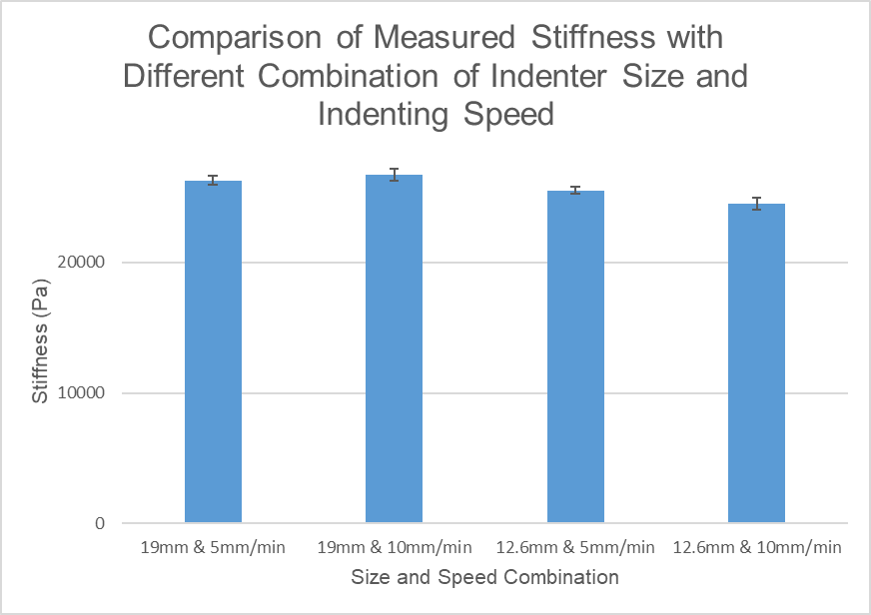}
    \caption{Pure PAAm with \SI{3}{\percent} crosslinker ratio sample is shown here as an example. Ball bearing size is set to 19 mm or 12.6 mm (0.75 in or 0.5 in), compression speed is set to 5 mm/min or 10 mm/min. The invariance in stiffness results with indenter size and speed show that our testing procedure is reliable and can be assumed quasistatic.}
    \label{SM Figure 2}
\end{figure}

\clearpage

\begin{figure}
    \centering
    \includegraphics[width=0.5\linewidth]{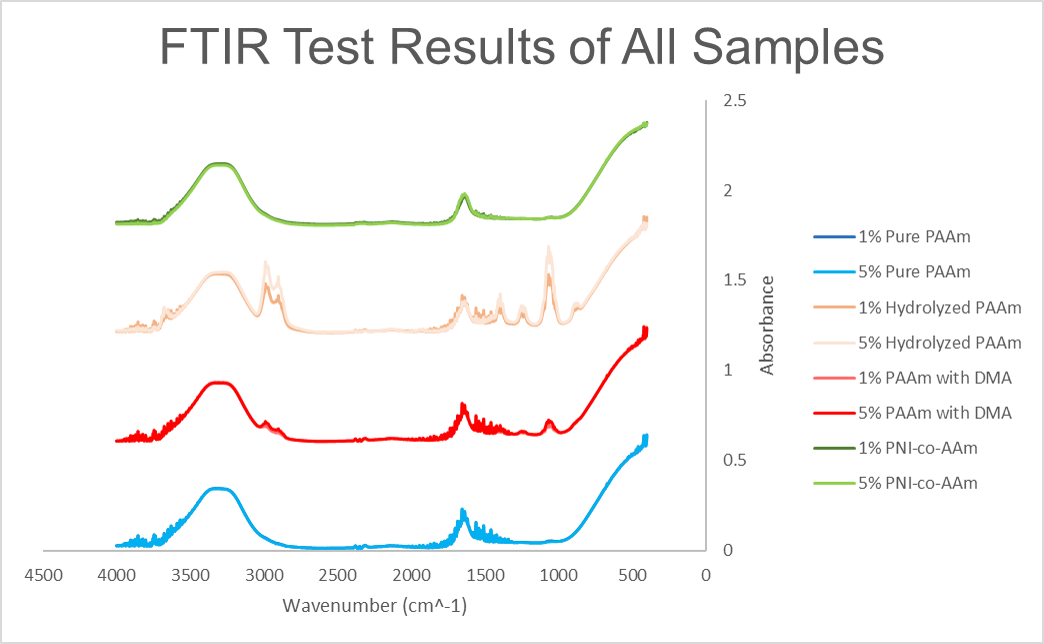}
    \caption{FTIR test results for samples from four hydrogel families. The results indicate that samples from the same hydrogel family only differ in polymer volume fraction. Plots have been shifted vertically by family for clarity.}
    \label{SM Figure 4}
\end{figure}

\clearpage

\begin{figure*}
    \centering
    \includegraphics[width=0.75\linewidth]{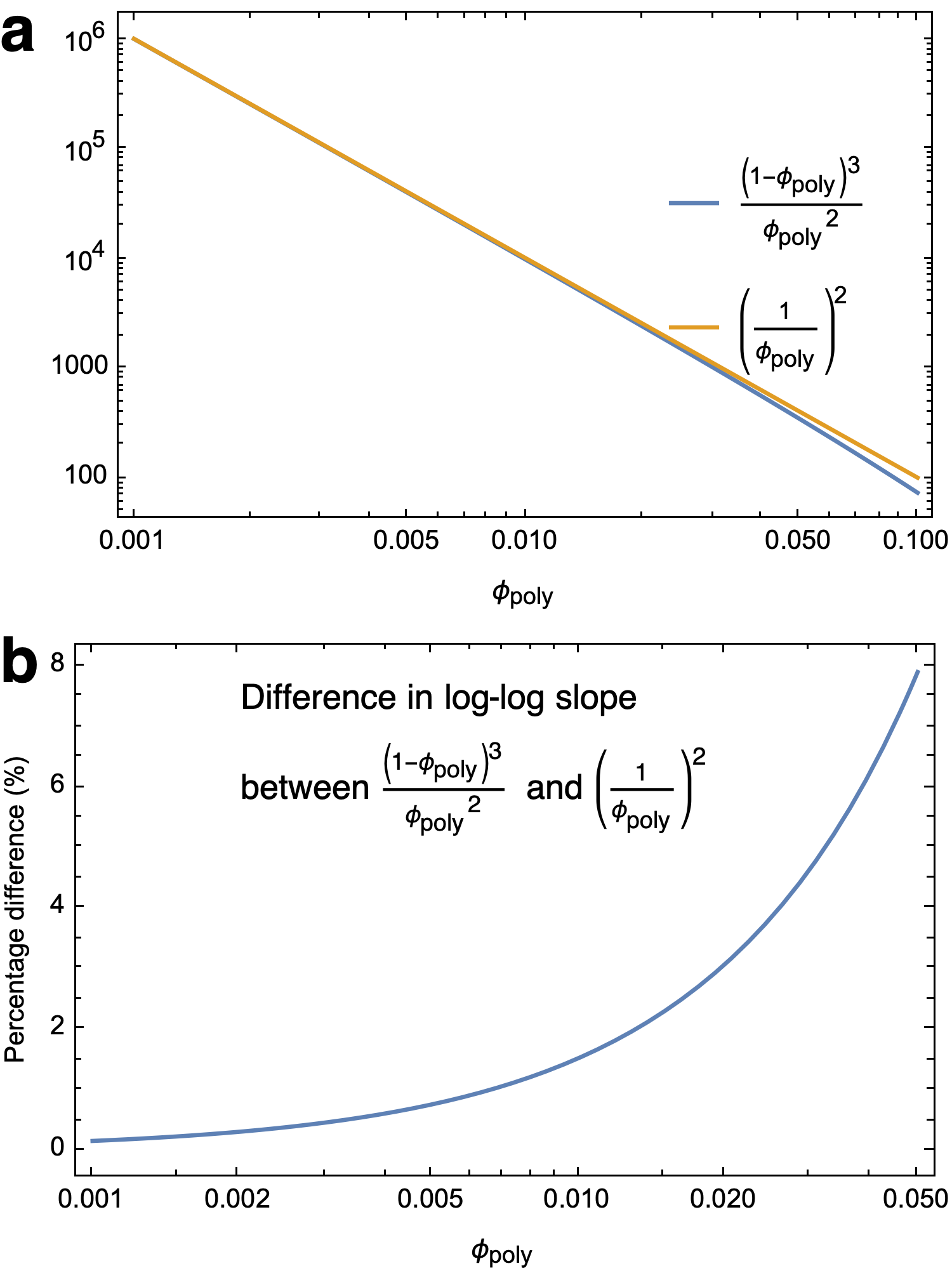}
    \caption{(a) For small $\phi_\text{poly}$ ($\phi_\text{poly}<0.05$) the permeability can be approximated as $\kappa \propto \phi_\text{poly}^{-2}$ with minimal error. (b) The calculation shows an error within \SI{8}{\percent} when $\phi_\text{poly}$ is equal or lower than \SI{0.05}.}
    \label{SM Figure 8}
\end{figure*}

\begin{figure*}
    \centering
    \includegraphics[width=0.75\textwidth]{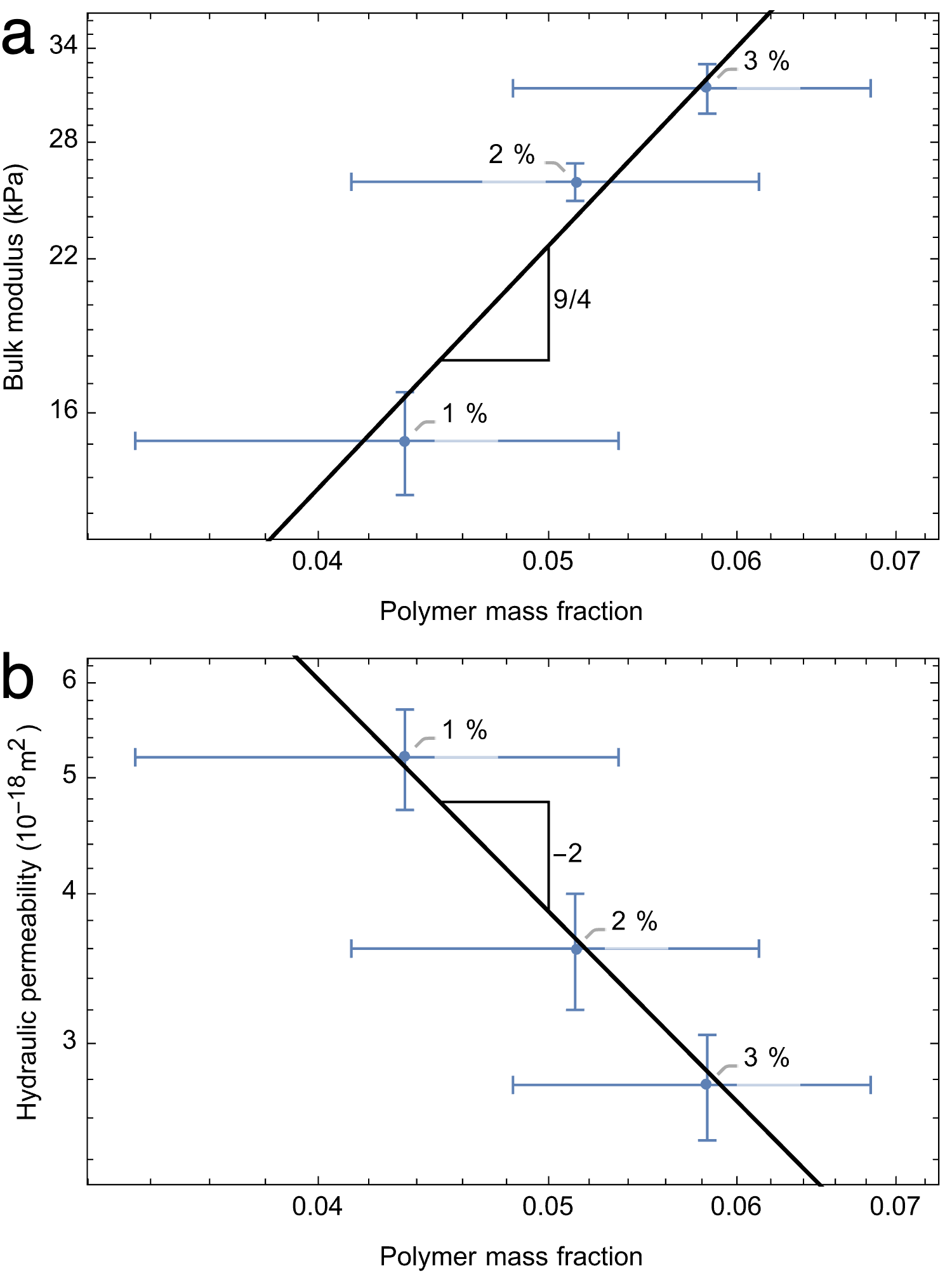}
    \caption{The stiffness and permeability of pure PAAm hydrogel samples with crosslinker ratios of 1, 2, and 3 \si{\percent} are plotted against the polymer mass fractions. The polymer mass fraction is proportional to the polymer volume fraction; thus, the stiffness should scale with the mass fraction by a $9/4$ power according to Eq. 2. (a) A log-log plot of bulk modulus versus polymer mass fraction confirms this $9/4$ scaling. Likewise, the permeability should scale with the mass fraction by a $-2$ power according to Eq. 7. (b) A log-log plot of hydraulic permeability versus polymer mass fraction confirms this $-2$ scaling. The polymer mass fractions were obtained by measuring the weight of samples dried for over one week under dry air flow and dividing by their weight after equilibrating in pure water for over one week.}
    \label{SM Figure 7}
\end{figure*}

\clearpage

\begin{table}
\begin{center}
\begin{tabular}{|l c c|} 
 \hline
Sample & Permeability, $\kappa$ ($10^{-18} \text{m}^2$) & Modulus, $K$ (kPa) \\ [0.5ex] 
 \hline\hline
Pure 0.5\% & $9.7 \pm 1.0$ & $7.0 \pm 0.6$\\ 
\hline
Pure 1\% & $5.2 \pm 0.5$ & $15.1 \pm 1.6$\\ 
\hline
Pure 2\% & $3.6 \pm 0.4$ & $25.8 \pm 1.0$\\ 
\hline
Pure 3\% & $2.8 \pm 0.3$ & $31.3 \pm 1.6$\\ 
\hline
Pure 5\% & $2.9 \pm 0.3$ & $31.0 \pm 0.9$\\ 
\hline
Hydro 1\% & $23.1 \pm 2.4$ & $7.6 \pm 0.7$\\ 
\hline
Hydro 2\% & $13.9 \pm 1.4$ & $14.4 \pm 1.7$\\ 
\hline
Hydro 3\% & $9.6 \pm 1.0$ & $18.4 \pm 0.7$\\ 
\hline
Hydro 5\% & $9.5 \pm 1.0$ & $20.9 \pm 1.2$\\ 
\hline
Hydro 6\% & $6.1 \pm 0.6$ & $35.6 \pm 0.9$\\ 
\hline
Hydro 7\% & $3.1 \pm 0.3$ & $85.7 \pm 2.4$\\ 
\hline
DMA 0.5\% & $12.6 \pm 1.3$ & $9.3 \pm 0.4$\\ 
\hline
DMA 1\% & $8.3 \pm 0.8$ & $14.6 \pm 0.3$\\ 
\hline
DMA 2\% & $5.6 \pm 0.6$ & $25.5 \pm 0.5$\\ 
\hline
DMA 3\% & $5.0 \pm 0.5$ & $31.8 \pm 0.4$\\ 
\hline
DMA 5\% & $5.4 \pm 0.6$ & $27.2 \pm 0.5$\\ 
\hline
NIP 0.5\% & $11.9 \pm 1.2$ & $8.0 \pm 0.7$\\ 
\hline
NIP 1\% & $6.0 \pm 0.6$ & $16.7 \pm 1.0$\\ 
\hline
NIP 2\% & $3.4 \pm 0.4$ & $32.9 \pm 0.6$\\ 
\hline
NIP 3\% & $2.9 \pm 0.3$ & $42.6 \pm 0.7$\\ 
\hline
NIP 5\% & $2.8 \pm 0.3$ & $43.3 \pm 0.5$\\
\hline
\end{tabular}
\end{center}
\caption{Stiffness, $K$, and permeability, $\kappa$, of all samples from four hydrogel families with varying crosslinker ratio.}
\end{table}

\clearpage

\begin{table}
\begin{center}
\begin{tabular}{|l c|} 
 \hline
 Hydrogel Family & Prefactor ($10^{-17}\text{m}^2\text{kPa}^{8/9}$) \\ [0.5ex] 
 \hline\hline
 Pure & $5.9 \pm 0.3$ \\
 DMA & $9.8 \pm 0.5$ \\
 Hydro & $14.5 \pm 0.7$ \\
 NIP & $7.8 \pm 0.4$ \\
 
\hline
\end{tabular}
\end{center}
\caption{Prefactors depend on the monomer size, solvent interaction and sphericity, thus, different hydrogel families have different values of prefactors. }
\end{table}

\bibliographystyle{ieeetr}
\providecommand{\noopsort}[1]{}\providecommand{\singleletter}[1]{#1}%


\begin{thebibliography}{10}

\bibitem{tan2019application}
H.-L. Tan, S.-Y. Teow, and J.~Pushpamalar, ``Application of metal
  nanoparticle--hydrogel composites in tissue regeneration,'' {\em
  Bioengineering}, vol.~6, no.~1, p.~17, 2019.

\bibitem{zhou2019highly}
Y.~Zhou, C.~Wan, Y.~Yang, H.~Yang, S.~Wang, Z.~Dai, K.~Ji, H.~Jiang, X.~Chen,
  and Y.~Long, ``Highly stretchable, elastic, and ionic conductive hydrogel for
  artificial soft electronics,'' {\em Advanced Functional Materials}, vol.~29,
  no.~1, p.~1806220, 2019.

\bibitem{subramani2020influence}
R.~Subramani, A.~Izquierdo-Alvarez, P.~Bhattacharya, M.~Meerts, P.~Moldenaers,
  H.~Ramon, and H.~Van~Oosterwyck, ``The influence of swelling on elastic
  properties of polyacrylamide hydrogels,'' {\em Frontiers in Materials},
  vol.~7, p.~212, 2020.

\bibitem{liu2020programmable}
K.~Liu, Y.~Zhang, H.~Cao, H.~Liu, Y.~Geng, W.~Yuan, J.~Zhou, Z.~L. Wu, G.~Shan,
  Y.~Bao, {\em et~al.}, ``Programmable reversible shape transformation of
  hydrogels based on transient structural anisotropy,'' {\em Advanced
  Materials}, vol.~32, no.~28, p.~2001693, 2020.

\bibitem{lin2016stretchable}
S.~Lin, H.~Yuk, T.~Zhang, G.~A. Parada, H.~Koo, C.~Yu, and X.~Zhao,
  ``Stretchable hydrogel electronics and devices,'' {\em Advanced Materials},
  vol.~28, no.~22, pp.~4497--4505, 2016.

\bibitem{jiang2011pva}
S.~Jiang, S.~Liu, and W.~Feng, ``Pva hydrogel properties for biomedical
  application,'' {\em Journal of the mechanical behavior of biomedical
  materials}, vol.~4, no.~7, pp.~1228--1233, 2011.

\bibitem{efron2007oxygen}
N.~Efron, P.~B. Morgan, I.~D. Cameron, N.~A. Brennan, and M.~Goodwin, ``Oxygen
  permeability and water content of silicone hydrogel contact lens materials,''
  {\em Optometry and Vision Science}, vol.~84, no.~4, pp.~E328--E337, 2007.

\bibitem{kim2008extended}
J.~Kim, A.~Conway, and A.~Chauhan, ``Extended delivery of ophthalmic drugs by
  silicone hydrogel contact lenses,'' {\em Biomaterials}, vol.~29, no.~14,
  pp.~2259--2269, 2008.

\bibitem{zhao2019novel}
L.~Zhao, P.~Wang, J.~Tian, J.~Wang, L.~Li, L.~Xu, Y.~Wang, X.~Fei, and Y.~Li,
  ``A novel composite hydrogel for solar evaporation enhancement at air-water
  interface,'' {\em Science of the total environment}, vol.~668, pp.~153--160,
  2019.

\bibitem{sinha2019advances}
V.~Sinha and S.~Chakma, ``Advances in the preparation of hydrogel for
  wastewater treatment: A concise review,'' {\em Journal of Environmental
  Chemical Engineering}, vol.~7, no.~5, p.~103295, 2019.

\bibitem{lu2021high}
H.~Lu, W.~Shi, F.~Zhao, W.~Zhang, P.~Zhang, C.~Zhao, and G.~Yu, ``High-yield
  and low-cost solar water purification via hydrogel-based membrane
  distillation,'' {\em Advanced Functional Materials}, vol.~31, no.~19,
  p.~2101036, 2021.

\bibitem{van2018hydrogel}
V.~Van~Tran, D.~Park, and Y.-C. Lee, ``Hydrogel applications for adsorption of
  contaminants in water and wastewater treatment,'' {\em Environmental Science
  and Pollution Research}, vol.~25, no.~25, pp.~24569--24599, 2018.

\bibitem{jones2008characterization}
D.~S. Jones, C.~P. Lorimer, C.~P. McCoy, and S.~P. Gorman, ``Characterization
  of the physicochemical, antimicrobial, and drug release properties of
  thermoresponsive hydrogel copolymers designed for medical device
  applications,'' {\em Journal of Biomedical Materials Research Part B: Applied
  Biomaterials: An Official Journal of The Society for Biomaterials, The
  Japanese Society for Biomaterials, and The Australian Society for
  Biomaterials and the Korean Society for Biomaterials}, vol.~85, no.~2,
  pp.~417--426, 2008.

\bibitem{lin2015influence}
S.~Lin and L.~Gu, ``Influence of crosslink density and stiffness on mechanical
  properties of type i collagen gel,'' {\em Materials}, vol.~8, no.~2,
  pp.~551--560, 2015.

\bibitem{pilipchuk2013influence}
S.~P. Pilipchuk, M.~K. Vaicik, J.~C. Larson, E.~Gazyakan, M.-H. Cheng, and
  E.~M. Brey, ``Influence of crosslinking on the stiffness and degradation of
  dermis-derived hydrogels,'' {\em Journal of Biomedical Materials Research
  Part A}, vol.~101, no.~10, pp.~2883--2895, 2013.

\bibitem{gao2021scaling}
Y.~Gao, N.~K. Chai, N.~Garakani, S.~S. Datta, and H.~J. Cho, ``Scaling laws to
  predict humidity-induced swelling and stiffness in hydrogels,'' {\em Soft
  Matter}, vol.~17, no.~43, pp.~9893--9900, 2021.

\bibitem{ju2010characterization}
H.~Ju, A.~C. Sagle, B.~D. Freeman, J.~I. Mardel, and A.~J. Hill,
  ``Characterization of sodium chloride and water transport in crosslinked poly
  (ethylene oxide) hydrogels,'' {\em Journal of Membrane Science}, vol.~358,
  no.~1-2, pp.~131--141, 2010.

\bibitem{peng2012ion}
C.-C. Peng and A.~Chauhan, ``Ion transport in silicone hydrogel contact
  lenses,'' {\em Journal of membrane science}, vol.~399, pp.~95--105, 2012.

\bibitem{pozuelo2014oxygen}
J.~Pozuelo, V.~Compa{\~n}, J.~M. Gonz{\'a}lez-M{\'e}ijome, M.~Gonz{\'a}lez, and
  S.~Moll{\'a}, ``Oxygen and ionic transport in hydrogel and silicone-hydrogel
  contact lens materials: An experimental and theoretical study,'' {\em Journal
  of membrane science}, vol.~452, pp.~62--72, 2014.

\bibitem{liu2008gas}
L.~Liu, A.~Chakma, and X.~Feng, ``Gas permeation through water-swollen hydrogel
  membranes,'' {\em Journal of Membrane Science}, vol.~310, no.~1-2,
  pp.~66--75, 2008.

\bibitem{matsuyama1997analysis}
H.~Matsuyama, M.~Teramoto, and H.~Urano, ``Analysis of solute diffusion in poly
  (vinyl alcohol) hydrogel membrane,'' {\em Journal of membrane science},
  vol.~126, no.~1, pp.~151--160, 1997.

\bibitem{yazdi2020hydrogel}
M.~K. Yazdi, V.~Vatanpour, A.~Taghizadeh, M.~Taghizadeh, M.~R. Ganjali, M.~T.
  Munir, S.~Habibzadeh, M.~R. Saeb, and M.~Ghaedi, ``Hydrogel membranes: A
  review,'' {\em Materials Science and Engineering: C}, vol.~114, p.~111023,
  2020.

\bibitem{sagle2009peg}
A.~C. Sagle, H.~Ju, B.~D. Freeman, and M.~M. Sharma, ``Peg-based hydrogel
  membrane coatings,'' {\em Polymer}, vol.~50, no.~3, pp.~756--766, 2009.

\bibitem{tavera2018characterization}
M.~J. Tavera-Quiroz, J.~J.~F. D{\'\i}az, and A.~Pinotti, ``Characterization of
  methylcellulose based hydrogels by using citric acid as a crosslinking
  agent,'' {\em International Journal of Applied Engineering Research},
  vol.~13, no.~17, pp.~13302--13307, 2018.

\bibitem{de1979scaling}
P.-G. De~Gennes, {\em Scaling concepts in polymer physics}.
\newblock Cornell university press, 1979.

\bibitem{vandersman2015biopolymer}
R.~{van der Sman}, ``Biopolymer gel swelling analysed with scaling laws and
  flory–rehner theory,'' {\em Food Hydrocolloids}, vol.~48, pp.~94--101,
  2015.

\bibitem{ZRINYI19871139}
M.~Zrinyi and F.~Horkay, ``On the elastic modulus of swollen gels,'' {\em
  Polymer}, vol.~28, no.~7, pp.~1139--1143, 1987.

\bibitem{obhukhov}
S.~P. Obukhov, M.~Rubinstein, and R.~H. Colby, ``Network modulus and
  superelasticity,'' {\em Macromolecules}, vol.~27, no.~12, pp.~3191--3198,
  1994.

\bibitem{bellpeppas}
C.~L. Bell and N.~A. Peppas, ``Biomedical membranes from hydrogels and
  interpolymer complexes,'' in {\em Biopolymers II} (N.~A. Peppas and R.~S.
  Langer, eds.), (Berlin, Heidelberg), pp.~125--175, Springer Berlin
  Heidelberg, 1995.

\bibitem{da2007effect}
R.~da~Silva and M.~G. de~Oliveira, ``Effect of the cross-linking degree on the
  morphology of poly (nipaam-co-aac) hydrogels,'' {\em Polymer}, vol.~48,
  no.~14, pp.~4114--4122, 2007.

\bibitem{ben2018comparative}
N.~E. Ben~Ammar, T.~Saied, M.~Barbouche, F.~Hosni, A.~H. Hamzaoui, and
  M.~{\c{S}}en, ``A comparative study between three different methods of
  hydrogel network characterization: Effect of composition on the crosslinking
  properties using sol--gel, rheological and mechanical analyses,'' {\em
  Polymer Bulletin}, vol.~75, no.~9, pp.~3825--3841, 2018.

\bibitem{collins2011morphology}
M.~N. Collins and C.~Birkinshaw, ``Morphology of crosslinked hyaluronic acid
  porous hydrogels,'' {\em Journal of Applied Polymer Science}, vol.~120,
  no.~2, pp.~1040--1049, 2011.

\bibitem{bryant2004crosslinking}
S.~J. Bryant, K.~S. Anseth, D.~A. Lee, and D.~L. Bader, ``Crosslinking density
  influences the morphology of chondrocytes photoencapsulated in peg hydrogels
  during the application of compressive strain,'' {\em Journal of Orthopaedic
  Research}, vol.~22, no.~5, pp.~1143--1149, 2004.

\bibitem{li2012experimental}
J.~Li, Y.~Hu, J.~J. Vlassak, and Z.~Suo, ``Experimental determination of
  equations of state for ideal elastomeric gels,'' {\em Soft Matter}, vol.~8,
  no.~31, pp.~8121--8128, 2012.

\bibitem{saber2012uv}
S.~Saber-Samandari, M.~Gazi, and E.~Yilmaz, ``Uv-induced synthesis of
  chitosan-g-polyacrylamide semi-ipn superabsorbent hydrogels,'' {\em Polymer
  bulletin}, vol.~68, no.~6, pp.~1623--1639, 2012.

\bibitem{aalaie2007preparation}
J.~Aalaie and A.~Rahmatpour, ``Preparation and swelling behavior of partially
  hydrolyzed polyacrylamide nanocomposite hydrogels in electrolyte solutions,''
  {\em Journal of Macromolecular Science, Part B}, vol.~47, no.~1, pp.~98--108,
  2007.

\bibitem{skelton2013biomimetic}
S.~Skelton, M.~Bostwick, K.~O'Connor, S.~Konst, S.~Casey, and B.~P. Lee,
  ``Biomimetic adhesive containing nanocomposite hydrogel with enhanced
  materials properties,'' {\em Soft Matter}, vol.~9, no.~14, pp.~3825--3833,
  2013.

\bibitem{kuru2007preparation}
E.~A. Kuru, N.~Orakdogen, and O.~Okay, ``Preparation of homogeneous
  polyacrylamide hydrogels by free-radical crosslinking copolymerization,''
  {\em European polymer journal}, vol.~43, no.~7, pp.~2913--2921, 2007.

\bibitem{manjula2013preparation}
B.~Manjula, K.~Varaprasad, R.~Sadiku, and K.~M. Raju, ``Preparation and
  characterization of sodium alginate--based hydrogels and their in vitro
  release studies,'' {\em Advances in polymer technology}, vol.~32, no.~2,
  2013.

\bibitem{yang2020preparation}
L.~Yang, X.~Fan, J.~Zhang, and J.~Ju, ``Preparation and characterization of
  thermoresponsive poly (n-isopropylacrylamide) for cell culture
  applications,'' {\em Polymers}, vol.~12, no.~2, p.~389, 2020.

\bibitem{louf2021underpressure}
J.-F. Louf, N.~B. Lu, M.~G. O’Connell, H.~J. Cho, and S.~S. Datta, ``Under
  pressure: Hydrogel swelling in a granular medium,'' {\em Science Advances},
  vol.~7, no.~7, p.~eabd2711, 2021.

\bibitem{nakao1979characteristics}
S.-I. Nakao, T.~Nomura, and S.~Kimura, ``Characteristics of macromolecular gel
  layer formed on ultrafiltration tubular membrane,'' {\em AIChE Journal},
  vol.~25, no.~4, pp.~615--622, 1979.

\bibitem{zaidi2005experimental}
S.~Zaidi and A.~Kumar, ``Experimental analysis of a gel layer in dead-end
  ultrafiltration of a silica suspension,'' {\em Desalination}, vol.~172,
  no.~2, pp.~107--117, 2005.

\bibitem{iritani2006compression}
E.~Iritani, N.~Katagiri, K.~Yamaguchi, and J.-H. Cho,
  ``Compression-permeability properties of compressed bed of superabsorbent
  hydrogel particles,'' {\em Drying technology}, vol.~24, no.~10,
  pp.~1243--1249, 2006.

\bibitem{isobe2018poroelasticity}
N.~Isobe, S.~Kimura, M.~Wada, and S.~Deguchi, ``Poroelasticity of cellulose
  hydrogel,'' {\em Journal of the Taiwan Institute of Chemical Engineers},
  vol.~92, pp.~118--122, 2018.

\bibitem{zhang2006surprising}
W.~Zhang, Y.~Liu, M.~Zhu, Y.~Zhang, X.~Liu, H.~Yu, Y.~Jiang, Y.~Chen,
  D.~Kuckling, and H.-J.~P. Adler, ``Surprising conversion of nanocomposite
  hydrogels with high mechanical strength by posttreatment: From a low swelling
  ratio to an ultrahigh swelling ratio,'' {\em Journal of Polymer Science Part
  A: Polymer Chemistry}, vol.~44, no.~22, pp.~6640--6645, 2006.

\bibitem{lv2019enhanced}
Q.~Lv, M.~Wu, and Y.~Shen, ``Enhanced swelling ratio and water retention
  capacity for novel super-absorbent hydrogel,'' {\em Colloids and Surfaces A:
  Physicochemical and Engineering Aspects}, vol.~583, p.~123972, 2019.

\bibitem{kim2003swelling}
S.~J. Kim, S.~J. Park, and S.~I. Kim, ``Swelling behavior of interpenetrating
  polymer network hydrogels composed of poly (vinyl alcohol) and chitosan,''
  {\em Reactive and Functional Polymers}, vol.~55, no.~1, pp.~53--59, 2003.

\bibitem{kumar2010synthesis}
A.~Kumar, M.~Pandey, M.~Koshy, and S.~A. Saraf, ``Synthesis of fast swelling
  superporous hydrogel: effect of concentration of crosslinker and acdisol on
  swelling ratio and mechanical strength,'' {\em Int. J Drug Deliv}, vol.~2,
  pp.~135--140, 2010.

\bibitem{kim2021fracture}
J.~Kim, G.~Zhang, M.~Shi, and Z.~Suo, ``Fracture, fatigue, and friction of
  polymers in which entanglements greatly outnumber cross-links,'' {\em
  Science}, vol.~374, no.~6564, pp.~212--216, 2021.

\bibitem{chen2019entanglement}
K.~Chen, Y.~Feng, Y.~Zhang, L.~Yu, X.~Hao, F.~Shao, Z.~Dou, C.~An, Z.~Zhuang,
  Y.~Luo, {\em et~al.}, ``Entanglement-driven adhesion, self-healing, and high
  stretchability of double-network peg-based hydrogels,'' {\em ACS applied
  materials \& interfaces}, vol.~11, no.~40, pp.~36458--36468, 2019.

\bibitem{gong2010double}
J.~P. Gong, ``Why are double network hydrogels so tough?,'' {\em Soft Matter},
  vol.~6, no.~12, pp.~2583--2590, 2010.

\bibitem{du2020highly}
R.~Du, Z.~Xu, C.~Zhu, Y.~Jiang, H.~Yan, H.-C. Wu, O.~Vardoulis, Y.~Cai, X.~Zhu,
  Z.~Bao, {\em et~al.}, ``A highly stretchable and self-healing supramolecular
  elastomer based on sliding crosslinks and hydrogen bonds,'' {\em Advanced
  Functional Materials}, vol.~30, no.~7, p.~1907139, 2020.

\bibitem{shi2020ultrastrong}
C.-Y. Shi, Q.~Zhang, C.-Y. Yu, S.-J. Rao, S.~Yang, H.~Tian, and D.-H. Qu, ``An
  ultrastrong and highly stretchable polyurethane elastomer enabled by a
  zipper-like ring-sliding effect,'' {\em Advanced Materials}, vol.~32, no.~23,
  p.~2000345, 2020.

\end{thebibliography}





\end{document}